\documentstyle[preprint,aps]{revtex}
\begin{document}
\draft
\preprint{}
\title{Quantum Hall effect in single wide quantum wells}
\author{M. Abolfath$^{1,3}$
\footnote{email: abolfath@gibbs.physics.indiana.edu},
L. Belkhir$^{1,2}$, and N. Nafari$^3$}
\address{$^1$Department of Physics, Indiana University, Bloomington, 
Indiana 47405}
\address{$^2$Xerox Corporation, 800 Phillips Road, 147-59B, Webster, NY 14580}
\address{$^3$Center for theoretical Physics and Mathematics, Atomic Energy
Organization of Iran, P.O.Box 11365-8486, Tehran, Iran}
\date{\today}
\maketitle
\begin{abstract}
We study the quantum Hall states in the lowest Landau 
level for a single wide quantum well. 
Due to a separation of charges to opposite sides of the well, a single
wide well can be modelled as an effective two level system.
We provide numerical evidence of the existence of a phase transition
from an incompressible to a compressible state as the electron
density is increased for specific well width. 
Our numerical results show a critical electron density which 
depends on well width,
beyond which a transition incompressible double layer quantum Hall state
to a mono-layer compressible state occurs. 
We also calculate the related phase boundary corresponding to destruction of 
the collective mode energy gap.
We show that the effective tunneling term and the interlayer separation
are both renormalised by the strong magnetic field.
We also exploite the local density functional techniques in the presence of
strong magnetic field at $\nu=1$ to calculate renormalized $\Delta_{SAS}$.
The numerical results shows good agreement between 
many-body calculations and local density functional techniques
in the presence of a strong magnetic field at $\nu=1$.
we also discuss implications of this work on the $\nu=1/2$ incompressible
state observed in SWQW.
\end{abstract}

\section{Introduction}
Recent interest in the properties of high-mobility quasi three-dimensional  
electron systems \cite{Suen91,Suen92} and double quantum well structures
(DQWs) in strong magnetic fields \cite{Boebinger90,Eisen-92} 
which are quite different in their structural details, has led to  
discovery of a novel forms of the fractional quantum Hall effect.
In the study by Eisenstein {\em et al.} \cite{Eisen-92}, 
the sample is a traditional symmetric double quantum well
(DQWs) structure. 
The envelope wave function in the middle of the sample is small
because there is a large barrier between the two wells,
whereas in the work of Suen {\em et al.} \cite{Suen91,Suen92}
the samples are actually a single wide quantum wells (SWQWs).
The self-consistent
electric field arising from the presence of the electrons themselves, 
splits the well into two spatially separated electron layers in the 
z-direction, effectively creating a double layer structure,
having a small self-consistent barrier with a finite envelope 
function in the middle.

Based on the fact that these samples are single wide quantum wells,
it has been argued by some authors that the quantum Hall mechanism 
of these systems is different from DQWs.
However, it has been claimed by 
Song He {\em et al.}~\cite{He-93}
that the result of finite size exact diagonalization calculation 
disagrees with this viewpoint and the bare $\Delta_{SAS}$,
the difference between the second and first subband energy
in zero magnetic field, which is reported by Ref. \cite{Suen92}
is not the relevant parameter determining the nature of the ground state in a
strongly correlated system. 
One may expect that the electronic structure of a SWQW is 
affected by a strong magnetic
field and therefore it changes the $\Delta_{SAS}$. In other words
$\Delta_{SAS}$ is renormalized by a strong magnetic field.
It has been claimed that the renormalized
$\Delta^B_{SAS}$ 
is small enough for the Halperin 
state \cite{Halperin83}, 
to be the ground state \cite{He-93}.
In this article we report the numerical calculation of the
renormalized $\Delta^B_{SAS}$ for 
different realistic models of the samples in Ref. \cite{Suen92}.
Our analytical expression for renormilized tunneling at $\nu=1$ 
is equivalent with the results of Hartree-Fock approximation (HFA).
We also implicate the renormilized tunneling's difinition for
fractional filling factors e.g. $\nu=1/2$.
Our first model is a many body Hamiltonian calculation with 
electron-electron, $isospin-potential$ interactions, 
where one may think of the layer index as an isospin index
and the second is based on the local density functional approximation (LDFA)
in the presence of a strong magnetic field. We also use LDFA to evaluate
the bare and renormalized tunneling at filling factor $\nu=1$.
There are some special cases in which SWQWs may effectively described
by DQWs, i.e. high areal electron density and large well width. 
In this article
DQW is the jargon for SWQW at high density or large width where the
overlap of wave functions is negligible in the middle of the well.
In these limits
the renormalized $\Delta^B_{SAS}$, which is the difference energy between 
the first and second subbands, in the presence of a strong perpendicular
magnetic field, is small enough and the quantum Hall ground states may
be described by Halperin's wave functions. In the past few years
SWQW's have been approximated by DQW, but there are significant correction
to the DQW Hamiltonian due to overlap of electrons wavefunction which are
localized in the SWQWs edges. In this article, we show that
these corrections become more important either
by decreasing the electron density or decreasing the well width.
We also show that the 
two-component quantum Hall ground state at $\nu=1$ which is stabilized by
a small $\Delta^B_{SAS}$ at high $N_s$ may evolve continuously to a
one-component $normal \; state$ at large $\Delta^B_{SAS}$ and low $N_s$
where we call a state of the 2DES with 
no quantum Hall effects by normal state.

Throughout this paper we neglect any explicit consideration of electron
spin, assuming that the magnetic field $B$ is always high enough to
totally spin polarize the system due to a large Zeeman splitting. 
We also ignore any effect of
Landau level mixing and higher subbands, working exclusively in the lowest
spin-split Landau level of the lowest subband. These assumptions are 
consistent with all the specific experiments to be disscused in this paper.
It is obvious that the DQW and SWQW have some common properties which
bring them into the same class. 
For the sake of simplicity, we use the term of 
$bilayer \; electron \; systems$
(BLES) to unified DQW and SWQW as long as we concern about their 
common properties.
There are  two energy scales
associated with a BLES in the presence of strong magnetic field,
$\Delta^B_{SAS}$ and the many-body Coulomb interaction between electrons
$e^2/\epsilon_0\ell_0^2$ where $\epsilon_0$ is dielectric constant
and $\ell_0=\sqrt{\hbar c/e B}$ is the magnetic length corresponding to
cyclotron orbit radius with electron cyclotron energy $\hbar \omega_c$,
where $\omega_c=eB/m^\ast c$ is the cyclotron frequency. 
Competition between these two energies 
leads to a rich phase diagram.
In a single layer two-dimensional electron system (2DES),
the integer quantum Hall effect (IQHE) is related to gaps in the
single particle density of states produced by the electron cyclotron
energy ($\hbar \omega_c$), while in the fractional quantum Hall effect 
(FQHE), the gap in the excitation spectrum is the result of the
electrons' Coulomb interaction. The quantum Hall plateau for 
BLES is controlled by the total filling factor ($\nu_{tot}$)
which is the sum of the filling factors in each layer. Therefore
the odd-IQHE of BLES corresponds to even denominator
filling factor for each layer, for example $\nu_{tot}=1$ is produced
by $\nu=1/2$ in each layer, even though $\nu=1/2$ in a single layer does
not yield a plateau.
Observation of odd-IQHE is due to correlation between electrons
in different layers which is destroyed if they are uncorrelated by increasing
the layer distance or decreasing $\Delta^B_{SAS}$, 
although for DQW with small enough separations the IQHE survives even
in the limit $\Delta^B_{SAS} \rightarrow 0$, due to a 
spontanous broken symmetry.
This phase transtion
has been observed for DQW by Boebinger {\em et al.}~\cite{Boebinger90} and
explained theoretically by MacDonald {\em et al.}~\cite{Mac90}. 
This effect has also been observed for SWQW by Suen 
{\em et al.}~\cite{Suen91}. In this article we present a theoretical
study of SWQW incompressible to compressible phase transitions at filling
factor $\nu=1$. 

In Section \ref{sec2}, we obtain the many body Hamiltonian of SWQWs
in the absence of magnetic field. We study the effect of strong
magnetic field  
in Section \ref{sec3} by projection of Hamiltonian onto 
the lowest Landaue level's Hilbert space and comparing with the 
Hartree-Fock approximation at $\nu=1$.
In Section \ref{sec4}, we use 
local density functional approximation 
in the presence of perpendicular strong magnetic field  
to find SWQWs band structure to 
evaluate the renormalized
$\Delta_{SAS}$ and comparing with other results.
In Section \ref{sec5},
the collective modes in quantum Hall state is studied. We present
a phase diagram for quantum Hall compressible-incompressible phase 
transition and
comparing with experimental results.


\section{Many body Hamiltonian of Single Wide Quantum Well
in zero magnetic field} 
\label{sec2}

In a SWQW the electrons are confined in the $x-y$
plane by an external potential barrier and therfore the 
energy spectrum becomes quantized into electric subbands. The number
of filled subbands is a function of the areal density, $N_s$. In experimental
work at zero magnetic field, the first two subbands are typically 
filled \cite{Suen91,Suen92}.
Neglecting high subbands, the subband degree of freedom
SWQWs effectively reduces to a two level system. This system may mapped
to a 2D electron system with an $isospin$ degree of freedom.
In the isospin language the isospin states,
$up \; (\uparrow)$ and $down \;  (\downarrow)$ refer to first and 
second subbands respectively.
In the absence of perpindicular electric field bias, the system is balanced
and the up (down) states are reflection
symmetric (antisymmetric) states respectively,  
with bare eigenenergy difference $\Delta^0_{SAS}$.
For the sake of simplicity we define $\Delta_{SAS}$ as the eigenenergy 
differece between the two subbands in both balanced and imbalanced cases.
The general second quantized many body hamiltonian in this picture is

\begin{equation}
\hat{H}=\hat{T}+\hat{V} \; ,
\label{1.0}
\end{equation}
where $\hat{T}$, in analogy to the double well case is refered to as
the tunneling energy
\begin{equation}
\hat{T}= -\frac{\Delta^0_{SAS}}{2} \sum_{{\bf p}}
( \; \hat{C}_{{\bf p}}^{\dagger\uparrow}\hat{C}_{{\bf p}}^{\uparrow}-
\hat{C}_{{\bf p}}^{\dagger\downarrow}\hat{C}_{{\bf p}}^{\downarrow} \;) \; . 
\label{1.1}
\end{equation}
Here $\Delta^0_{SAS}$ is the subband energy difference of noninteracting 
electron system in the SWQW.
The interaction part of the many-body Hamiltonian employed in our studies  
can be written in zero magnetic field as

\begin{equation}
\hat{V}=\frac{1}{2}\sum_{{\bf p},{\bf p}^{\prime},{\bf q}}\sum_{\{\sigma\}}
V^{\sigma^{\prime}_{1}\sigma^{\prime}_{2}\sigma_{1}\sigma_{2}}_{{\bf q}}
\hat{C}^{\dagger\sigma^{\prime}_{1}}_{{\bf p}+{\bf q}}
\hat{C}^{\dagger\sigma^{\prime}_{2}}_{{\bf p}^{\prime}-{\bf q}}
\hat{C}^{\sigma_{2}}_{{\bf p}^{\prime}}\hat{C}^{\sigma_{1}}_{{\bf p}} \; .
\label{1.2}
\end{equation}
In the above equation
$V^{\sigma^{\prime}_{1}\sigma^{\prime}_{2}\sigma_{1}\sigma_{2}}_{{\bf q}}$ 
is the Fourier transform of the electron-electron Coulomb interaction

\begin{equation}
V^{\sigma^{\prime}_{1}\sigma^{\prime}_{2}\sigma_{1}\sigma_{2}}_{{\bf q}}=
\frac{2\pi e^2}{\epsilon_0 \; q} \int dz_{1}\int dz_{2} \;
\psi^{\ast\sigma^{\prime}_{1}}(z_{1})\psi^{\ast\sigma^{\prime}_{2}}(z_{2})
\psi^{\sigma_{1}}(z_{1})
\psi^{\sigma_{2}}(z_{2}) e^{- q|z_{1}-z_{2}|} \, ,
\label{1.3}
\end{equation}
where ${\bf q}$ is the in plane wavevector.
$\hat{C}^{\dagger\sigma}_{\bf p}(\hat{C}^{\sigma}_{\bf p})$ is the 
creation (anihilation) operator for 
electrons with 2D momentum ${\bf p}$ and isospin state $\sigma$ and obeys,

\begin{equation}
\{ \hat{C}^{\sigma}_{{\bf p}},
\hat{C}^{\dagger\sigma^{\prime}}_{{\bf p}^{\prime}} \}=
\delta_{{\bf p},{\bf p}^{\prime}} \, \delta^{\sigma,\sigma^{\prime}} \, .
\end{equation}

It is convenient to define the $4-$vector isospin density operator
\begin{eqnarray}
&\sigma^{\mu}_{\bf k} = 
\Bigl( \sigma^0_{\bf k} \; , \; \sigma^{a}_{\bf k} \Bigl) \; , \nonumber \\&
\sigma^a_{\bf k} = \Bigl(\sigma^x_{\bf k} \; ,\;
\sigma^y_{\bf k} \; ,\; \sigma^z_{\bf k} \Bigl) \, ,
\end{eqnarray}
where $\sigma^0$ is unit matrix and $\sigma^a$ is the $a$th component
of Pauli matrices

\begin{equation}
\sigma^{x}=\left(\begin{array}{cc}
0 & 1\\
1 & 0 \end{array}\right),\hskip .25in
\sigma^{y}=\left(\begin{array}{cc}
0 & -i\\
i & 0 \end{array}\right),\hskip .25in
\sigma^{z}=\left(\begin{array}{cc}
1 & 0\\
0 & -1 \end{array}\right).
\end{equation}
Here $\mu \in \{0,1,2,3\}$.
The zeroth component of $\sigma^\mu_{\bf k}$ is the scalar density 
operator and the rest is the vectorial isospin density.
One may write the $4-$vector isospin density operator in terms of
creation-annihilation operators
\begin{equation}
\hat{\sigma}^{\mu}_{\bf q}=\sum_{\bf p}
\left(
\hat{C}^{\dagger \uparrow}_{{\bf p}-{\bf q}} \hskip .1in 
\hat{C}^{\dagger \downarrow}_{{\bf p}-{\bf q}} 
\right)
\, \sigma^{\mu} \,
\left(\begin{array}{c}
\hat{C}^{\uparrow}_{{\bf p}} \\ \hat{C}^{\downarrow}_{{\bf p}} 
\end{array}\right) \, .
\label{1.6}
\end{equation}

The transformed hamiltonian Eq.(\ref{1.0}) with continous 
wave vector ${\bf q}$ simplifies to
\begin{equation}
\hat{H}=-\frac{\Delta_{SAS}}{2} \sigma^z({\bf q}=0)
+ \frac{1}{2} \int \frac{d^2 {\bf q}}{(2 \pi)^2}
\sigma_\mu(-{\bf q}) \,  V^{\mu\nu} ({\bf q}) \; \sigma_\nu({\bf q}) \, .
\label{1.7}
\end{equation}
The isospin potential $V^{\mu\nu} ({\bf q})$ in Eq.(\ref{1.7}) describes 
the electron-electron isospin interaction, for the sake of simplicity 
we call the $V^{\mu\nu} ({\bf q})$, the $isopotential$ matrix elements. They 
are linear combinations of the coulomb potential form factors in Eq.(\ref{1.3})
and are defined in Eq.(\ref{1.8}).

One finds after a little algebraic calculation,
$\Delta^0_{SAS}$ is renormalized by the exchange electron-electron
interaction 
\begin{equation}
\Delta_{SAS} \; = \; \Delta^0_{SAS} \; 
+ 2  \int \frac{d^2 {\bf q}}{(2 \pi)^2} \, V_{0z}({\bf q}).
\label{bare}
\end{equation}
The enhancent may be positive or negative, depending on the sign of 
integral upon $V_{0z}({\bf q})$. 
For example it is positive if the density is low
and negative at high density for a given well width. Therefore
for large (small) bare $\Delta^0_{SAS}$ the renormalized 
$\Delta_{SAS}$ is greater (lesser) than $\Delta^0_{SAS}$ respectively.
In the section \ref{sec4}, we will calculate $\Delta_{SAS}$ by
local density functional approximation (LDFA) where the effect
of local exchange-correlation interaction is included in 
$\Delta_{SAS}$. 
One may show that the results of LDFA and HFA
are close, hence the left hand side of Eq.(\ref{bare})
can be evaluated by LDFA. 

The non-zero isopotentials elements
in Eq.(\ref{1.7}) may be written in terms of the 
electron-electron interaction of matrix elements defined in
Eq.(\ref{1.3}).

\begin{eqnarray}
&V_{00}({\bf q})=\frac{1}{4}\bigl(
V^{\uparrow\uparrow}_{{\bf q}}+2V^{\uparrow\downarrow}_{{\bf q}}
+V^{\downarrow\downarrow}_{{\bf q}}\bigl) \, , \nonumber \\
&V_{zz}({\bf q})=\frac{1}{4}\bigl(
V^{\uparrow\uparrow}_{{\bf q}}-2V^{\uparrow\downarrow}_{{\bf q}}
+V^{\downarrow\downarrow}_{{\bf q}}\bigl) \, , \nonumber \\
&V_{0x}({\bf q})=V_{x0}({\bf q})=\frac{1}{2}
\bigl(C^{\uparrow\downarrow}_{{\bf q}}
+D^{\uparrow\downarrow}_{{\bf q}}\bigl) \, , \nonumber \\
&V_{0z}({\bf q})=V_{z0}({\bf q})=\frac{1}{4}
\bigl(V^{\uparrow\uparrow}_{{\bf q}}
-V^{\downarrow\downarrow}_{{\bf q}}\bigl) \, , \nonumber \\
&V_{xz}({\bf q})=V_{zx}({\bf q})=\frac{1}{2}
\bigl(C^{\uparrow\downarrow}_{{\bf q}}
-D^{\uparrow\downarrow}_{{\bf q}}\bigl) \, , \nonumber \\
&V_{xx}({\bf q})=B^{\uparrow\downarrow}_{{\bf q}} \, ,
\label{1.8}
\end{eqnarray}
Taking the advantage of symmetry properties of Eq.(\ref{1.3}),
and the fact that the subband eigenfunctions are real, we define

\begin{equation}
V^{\uparrow\uparrow}_{{\bf q}}
\equiv
V^{\uparrow\uparrow\uparrow\uparrow}_{{\bf q}} \; ,
\hskip .5in 
V^{\downarrow\downarrow}_{{\bf q}} 
\equiv
V^{\downarrow\downarrow\downarrow\downarrow}_{{\bf q}} \; ,
\end{equation}
\begin{equation}
V^{\uparrow\downarrow}_{{\bf q}} 
\equiv
V^{\uparrow\downarrow\uparrow\downarrow}_{{\bf q}}= 
V^{\downarrow\uparrow\downarrow\uparrow}_{{\bf q}} \; ,
\end{equation}
\begin{equation}
B^{\uparrow\downarrow}_{{\bf q}} 
\equiv
V^{\uparrow\uparrow\downarrow\downarrow}_{{\bf q}}=
V^{\downarrow\uparrow\uparrow\downarrow}_{{\bf q}}=
V^{\uparrow\downarrow\downarrow\uparrow}_{{\bf q}}=
V^{\downarrow\downarrow\uparrow\uparrow}_{{\bf q}} \; ,
\end{equation}
\begin{equation}
C^{\uparrow\downarrow}_{{\bf q}} 
\equiv
V^{\uparrow\uparrow\uparrow\downarrow}_{{\bf q}}=
V^{\uparrow\uparrow\downarrow\uparrow}_{{\bf q}}=
V^{\uparrow\downarrow\uparrow\uparrow}_{{\bf q}}=
V^{\downarrow\uparrow\uparrow\uparrow}_{{\bf q}} \; ,
\label{30.3}
\end{equation}
\begin{equation}
D^{\uparrow\downarrow}_{{\bf q}} 
\equiv
V^{\downarrow\downarrow\downarrow\uparrow}_{{\bf q}}=
V^{\downarrow\downarrow\uparrow\downarrow}_{{\bf q}}=
V^{\downarrow\uparrow\downarrow\downarrow}_{{\bf q}}=
V^{\uparrow\downarrow\downarrow\downarrow}_{{\bf q}} \; .
\label{30.0}
\end{equation}
Note that the $y$ component of isopotential are zero.
In the next section we will see that these elements are still
zero in the presence of the prependicular magnetic field.
However, it has been found
that \cite{Ramin} the $y$ component are not zero due to
tilted magnetic field effect
and therefore all of isopotential elements
in Hamiltonian are nonzero. 
One may find by inversion symmetry of the Hamiltonian 
that both $C^{\uparrow\downarrow}_{{\bf q}}$
and $D^{\uparrow\downarrow}_{{\bf q}}$ are zero unless the system is 
imbalanced by applying a perpendicular external electric field.

Another extreme case may happen in the high density balanced regime,
shown in Fig.(\ref{fig0}). The overlap integral between the 
wavefunctions in the left and right is negligible 
and therefore we may show that,
$|\psi^{\uparrow}(z)|=|\psi^{\downarrow}(z)|$. 
In this regime one 
finds the only nonzero matrix elements are $B^{\uparrow\downarrow}_{{\bf q}}$, 
and
\begin{eqnarray}
V^{\uparrow\uparrow}_{{\bf q}}=V^{\downarrow\downarrow}_{{\bf q}}=
V^{\uparrow\downarrow}_{{\bf q}}=V^{\downarrow\uparrow}_{{\bf q}} \; .
\label{1.10}
\end{eqnarray}
In Fig.(\ref{fig01}), $V_{xx}({\bf q})$ and $V_{zz}({\bf q})$ is
shown for two different areal densities. At large areal density
$V_{zz}({\bf q})$ is negligible compared with $V_{xx}({\bf q})$ and 
the corresponding Hamiltonian is effectively the DQW Hamiltonian.
The non-zero isopotential matrix elements in this case are $V_{00}({\bf q})$
and $V_{xx}({\bf q})$ and may be obtained by
applying Eq.(\ref{1.10}) in Eq.(\ref{1.8}).
These matrix elements have been exploited before 
in the study of double quantum wells \cite{Mac90,Fertig89} where
the SWQW is effectively a DQW. Note that in this case $V_{0z}({\bf q})$
is nearly zero and the effect of many body exchange part of Hamiltonian
on $\Delta_{SAS}$ is negligible so that, $\Delta_{SAS}=\Delta^0_{SAS}$.
In the next section we will see that this is true even in the presence of
strong magnetic field. Therefore in the DQW or high density SWQW problems,
the effect of the many body part of Hamiltonian on $\Delta_{SAS}$
is zero. 

Therfore one may decompose the electron-electron interaction 
Hamiltonian Eq.(\ref{1.7}) into three terms

\begin{eqnarray}
\hat{V}=\hat{V}_{DL}+\hat{V}_{B}+\hat{V}_{U} .
\label{1.11}
\end{eqnarray}
where $\hat{V}_{DL}$, $\hat{V}_{B}$ and $\hat{V}_{U}$ are the double layer, 
balanced and unbalanced part of electron-electron interaction respectively

\begin{equation}
\hat{V}_{DL}=\frac{1}{2}\int \frac{d^2 {\bf q}}{(2 \pi)^2}
\left\{
V_{0}({\bf q})\hat{\rho}(-{\bf q})\hat{\rho}({\bf q}) 
+V_{x}({\bf q})\hat{\sigma}^x(-{\bf q})\hat{\sigma}^x({\bf q})\right\} \; ,
\end{equation}

\begin{eqnarray}
\hat{V}_{B}=\frac{1}{2}\int \frac{d^2 {\bf q}}{(2 \pi)^2}\left\{
V_{0z}({\bf q})[\hat{\rho}(-{\bf q})\hat{\sigma}^z({\bf q})
+\hat{\sigma}^z(-{\bf q})\hat{\rho}({\bf q})]
+V_z({\bf q})\hat{\sigma}^z(-{\bf q})\hat{\sigma}^z({\bf q}) \right\} 
\end{eqnarray}

\begin{equation}
\hat{V}_{U}=\frac{1}{2}\int \frac{d^2 {\bf q}}{(2 \pi)^2}\left\{
V_{0x}({\bf q})[\hat{\rho}(-{\bf q})\hat{\sigma}^x({\bf q})
+\hat{\rho}(-{\bf q})\hat{\sigma}^x({\bf q})]
+V_{xz}({\bf q})[\hat{\sigma}^x(-{\bf q})\hat{\sigma}^z({\bf q})+
\hat{\sigma}^z(-{\bf q})\hat{\sigma}^x({\bf q})]\right\} .
\label{1.12}
\end{equation}
$\hat{V}_{DL}$ has the same form as in a DQW problem.
The last two terms are corrections to the DQW problem, due to non-zero
electron density in the middle of the SWQW. 
These two terms are negligible 
when the density of electrons in the middle of the SWQW is small enough.

\section{many body hamiltonian of SWQWs in presence of
strong perpendicular magnetic field}
\label{sec3}

In this section we study the SWQW in the presence of strong magnetic field
where all electrons are located in the LLL. We exploit the techniques 
of projection onto the LLL which has been developed by Girvin and 
Jach \cite{Girvin-Jack}. 
We will show that this technique is equivalant with the Hartree-Fock 
approximation at filling factor $\nu=1$.

\subsection{Projection onto the lowest Landau level}
Taking the perpendicular component of the magnetic field to be strong, 
we restrict the Hilbert space to the lowest Landau level (LLL)
and exploit the LLL projection formalism
which was developed to study of collective 
excitations in the FQHE \cite{Girvin-Jack,GMP}. The interaction part of 
the Hamiltonian which describes the low-energy excitations of the system  
obtained in Eqs.($\ref{1.11}$) through ($\ref{1.12}$), may be projected 
onto the LLL by using

\begin{eqnarray}
&\overline{\rho_{-{\bf q}}\rho_{{\bf q}}}=\overline{\rho}_{-{\bf q}} 
\overline{\rho}_{{\bf q}}
-\overline{\rho}_{{\bf q}=0} \, e^{-q^2/2} , \nonumber \\&
\overline{\sigma^{a}_{-{\bf q}}\sigma^{b}_{{\bf q}}}=
\overline{\sigma}^{a}_{-{\bf q}}
\overline{\sigma}^{b}_{{\bf q}}-\bigl(i\epsilon^{abc}
\overline{\sigma}^c_{{\bf q}=0}+\delta^{ab}
\overline{\rho}_{{\bf q}=0}\bigl) \, e^{-q^2/2} , \nonumber \\&
\overline{\sigma^{\mu}_{-{\bf q}}\rho_{{\bf q}}}=
\overline{\sigma}^{\mu}_{-{\bf q}}\overline{\rho}_{{\bf q}}-
\overline{\sigma}^{\mu}_{{\bf q}=0} \, e^{-q^2/2} \; .
\label{2.1}
\end{eqnarray}
In the above equations $\overline{\rho}_{q}$ and $\overline{\sigma}^\mu_{q}$ 
are the projected total charge density
and $\mu$th component of isospin density operators respectively \cite{Moon}

\begin{mathletters}
\label{eq2.4}
\begin{equation}
\overline{\rho}({\bf q}) = \frac{1}{\sqrt{A}}\> 
\sum_{j=1}^N \overline{e^{-i{\bf q}\cdot
{\bf r}_j}} = \frac{1}{\sqrt{A}}\> \sum_{j=1}^N e^{-\frac{\vert
q\vert^2}{4}}\> \tau_q(j)
\label{eq2.4a}
\end{equation}
\begin{equation}
\overline{\sigma^\mu}({\bf q}) = \frac{1}{\sqrt{A}}\> \sum_{j=1}
^N \overline{e^{-i{\bf q}\cdot {\bf r}_j}}\> \sigma_j^\mu = 
\frac{1}{\sqrt{A}}\> \sum_{j=1}^N
e^{-\frac{\vert q\vert^2}{4}}\> \tau_q(j)\> \sigma_j^\mu ,
\label{eq2.4b}
\end{equation}
\end{mathletters}
\noindent where $q=\ell_0 (q_x + iq_y)$. 
The magnetic translation operator for the $j$th particle

\begin{equation}
\tau_q (j) = e^{-iq\frac{\partial}{\partial z_j} - \frac{i}{2}q^\ast z_j} \; ,
\label{eq3.90}
\end{equation}
is a unitary operator satisfying the closed Lie algebra

\begin{mathletters}
\label{eq3.100}
\begin{equation}
\tau_q\tau_k = \tau_{q+k}\> e^{\frac{i}{2}q\wedge k} ,
\label{eq3.100a}
\end{equation}
\begin{equation}
[\tau_q, \tau_k] = 2i\; \tau_{q+k}\> \sin{\frac{q\wedge k}{2}}
,\label{eq3.100b}
\end{equation}
\end{mathletters}

\noindent where 
$q\wedge k \equiv \ell^2_{0}({\bf q}\times {\bf k})\cdot {\hat{\bf z}}$. 

One may find the commutation relations between isospin density operators
by using Eq.(\ref{eq3.100}) 

\begin{mathletters}
\label{2.10}
\begin{eqnarray}
[\overline{\sigma}^{0}_{k_1}, \overline{\sigma}^\mu_{k_2}]
= (e^{k_1 k^{\ast}_2/2} - e^{k_2 k^{\ast}_1/2} ) 
   \overline{\sigma}^\mu_{k_1+k_2} \; , 
\label{2.10a}
\end{eqnarray}
\begin{eqnarray}
[\overline{\sigma}^a_{k_1}, \overline{\sigma}^b_{k_2}]
= (e^{k_1 k^{\ast}_2/2} - e^{k_2  k^{\ast}_1/2} )\; 
\overline{\sigma}^{0}_{k_1+k_2}\; \delta^{a b}
+(e^{k_1 k^{\ast}_2/2} + e^{k_2 k^{\ast}_1/2} )\; 
i \epsilon^{a b c}\;  \overline{\sigma}^{c}_{k_1+k_2} \; ,
\label{2.10b}
\end{eqnarray}
\end{mathletters}
where $k^{\ast}$ is the complex conjugate of $k$. Note that 
4-components are labeled by Greek indices and Latin indices
denotes spatial components.
The tunneling term is renormalized by the last linear term of Eq.(\ref{2.1}).  
The projected tunneling term is thus

\begin{equation}
\overline{T}=-\frac{\Delta^B_{SAS}}{2}\; \overline{\sigma}^z({\bf q}=0)\; ,
\end{equation}
where $\Delta^B_{SAS}$ is the renormalized energy difference 
between up and down isospin states. Due to the effect of the
strong magnetic field on the exchange energy, we have

\begin{equation}
\Delta^B_{SAS}=\Delta_{SAS} + 2 \int \frac{d^2 {\bf q}}{(2 \pi)^2}
V_{0z}({\bf q}) \, e^{-q^2 \ell^2_0/2} \; .
\label{2.1_1}
\end{equation}
One may obtain the analytical expression for renormilized tunneling, 
Eq.(\ref{2.1_1}), using the Hartree-Fock calculation where the many body
ground state wave function may be written as a single Slater determinant. 
A particular class of single Slater deteminant at $\nu=1$ in the Landaue 
guage which yields to Eq.(\ref{2.1_1}) can be written in the form \cite{Moon}
\begin{equation}
|\Psi\rangle=\prod_{X}(\hat{C}^\dagger_{X\uparrow}  \cos\frac{\theta(X)}{2}+
\hat{C}^\dagger_{X\downarrow}\sin\frac{\theta(X)}{2}e^{i\varphi(X)})|0\rangle
\label{eq10}
\end{equation}
where $|0\rangle$ is fermionic vacuum and 
$\hat{C}^\dagger_{X\uparrow,\downarrow}$ creates an electron in the 
Symmetric (Antiymmetric) or Left (Right) in orbit $\phi_{X}$, respectively. 
Taking the many body wave function Eq.(\ref{eq10}) 
and using the HFA, leads to Eq.(\ref{2.1_1}).
We may use Eq.(\ref{2.1_1}) for any fractional filling factors e.g. $\nu=1/2$
to evaluate the renormalized tunneling. 
Note that this generalization is just a simple 
extrapolation of Eq.(\ref{2.1_1}) to fractional filling factors.
As one may see from Eq.(\ref{2.1_1}), for a given well width and areal density,
$V_{0z}({\bf q})$ is given and $\Delta^B_{SAS}$ is just a monotonic
decreasing function of magnetic field 

\begin{equation}
\Delta^B_{SAS}(\nu)-\Delta_{SAS} \propto \mp \nu^{-1/2} \; , 
\label{2.11_2}
\end{equation}
where $\nu$ is the Landau level filling factor. 
The sign is specified by the integral in Eq.(\ref{2.1_1}),
which depends on the electron areal density.
Our numerical calculation shows that
in the domain of experimental densities, the sign is minues, i.e.
the renormalized $\Delta_{SAS}$ is less than the bare one.
It is instructive to write $\Delta^B_{SAS}(\nu)$ in terms of the
renormalized $\Delta_{SAS}$ at given filling factor i.e. $\nu=1$ 
and the bare $\Delta_{SAS}$
\begin{equation}
\Delta^B_{SAS}(\nu)= (1-\nu^{-1/2}) \Delta_{SAS}+
     \nu^{-1/2} \Delta^B_{SAS}(\nu=1) \; .
\label{2.12_2}
\end{equation}
We may use Eq.(\ref{2.12_2}) in order to find
$\Delta^B_{SAS}(\nu=1/2)$. 
To the best of our
knowledge there is no reliable measurement or calculation to find 
$\Delta^B{SAS}$ and interlayer seperation ($d$) at fractional filling factor.
Our numerical results for realistic samples
has been shown in Table I. 
According to thses results, we may conclude that the 
renormalized tunneling term at $\nu=1/2$
is large enough that 331 Halperin's wave function, which is exact
in the limit of vanishing tunneling, is not the best variational wavefunction
to describe the ground state. However, it is not obvious that
$\Delta^B_{SAS}(\nu \neq 1)$ is the relavant parameter to the FQH gap
corresponding to the FQHE at $\nu \neq 1$.

At the end of this section and for further calculations we list the
projected isopotentials onto LLL 

\begin{eqnarray}
\overline{V}=\overline{V}_{DL}+\overline{V}_{B}+\overline{V}_{U} \; ,
\label{2.2}
\end{eqnarray}

\begin{equation}
\overline{V}_{DL}=\frac{1}{2}\int \frac{d^2 {\bf q}}{(2 \pi)^2} \left\{
V_{0}({\bf q})\overline{\rho}(-{\bf q})\overline{\rho}({\bf q})+V_{x}({\bf q})
\overline{\sigma}^x(-{\bf q})\overline{\sigma}^x({\bf q}) \right\} \; ,
\end{equation}
\begin{equation}
\overline{V}_{B}=\frac{1}{2}\int \frac{d^2 {\bf q}}{(2 \pi)^2} \left\{
V_{0z}({\bf q})[\overline{\rho}(-{\bf q})\overline{\sigma}^z({\bf q})
+\overline{\sigma}^z(-{\bf q})\overline{\rho}({\bf q})]+V_z({\bf q})
\overline{\sigma}^z(-{\bf q})\overline{\sigma}^z({\bf q}) \right\} \; , 
\end{equation}
\begin{equation}
\overline{V}_{U}=\frac{1}{2}\int \frac{d^2 {\bf q}}{(2 \pi)^2} \left\{
V_{0x}({\bf q})[\overline{\rho}(-{\bf q})\overline{\sigma}^x({\bf q})
+\overline{\sigma}^x(-{\bf q})]\overline{\rho}({\bf q})
+V_{xz}({\bf q})[\overline{\sigma}^x(-{\bf q})\overline{\sigma}^z({\bf q})+
\overline{\sigma}^z(-{\bf q})\overline{\sigma}^x({\bf q})] \right\} \; .
\label{2.3}
\end{equation}
As we mentioned in the zero magnetic field case, the last two terms 
of Eq.($\ref{2.2}$), are negligible 
when the density of electrons in the middle of SWQWs is small enough,
and the system is effectively a DQW with renormalized $\Delta^B_{SAS}$
(which is nearly equal to the bare one) 
and hence one may expect that the many body quantum Hall ground state is 
described by the Halperin variational wave functions as it is in DQWs.

\section{Local density functional results for SWQWs
in strong magnetic field}
\label{sec4}

The system we study numerically in this section is a wide, single, 
GaAs quantum well in the
presence of a strong perpendicular magnetic field. 
We consider a slab of GaAs of thickness $W$ confined between two infinitely 
high barriers, which represent ${\rm Al_{x}Ga_{1-x}As}$. This system is doped 
by Si delta layers, separated symmetrically, by distance h from the GaAs and
which have ionized donor concentration $N_{s}$ per unit area.
We apply the local density functional approximation (LDFA) to find eigenvalues 
and eigenfunctions of the  many body system.
By this technique one may find the charge distribution 
function, eigenenergies and the effective potential, 
which are calculated by self-consistent Kohn-Sham equations
in a strong magnetic field, 
i.e. by solving the Poisson and Schr\"odinger equations simultaneously.
In the presence of a strong magnetic field, we restrict our attention 
to the lowest Landau level (LLL), for which the single-body wavefunctions, 
in symmetric gauge and polar coordinates are 

\begin{equation}
\varphi_{m}(r,\theta)=\frac{1}{\sqrt{2\pi\ell_{0}^{2}2^{m}m!}} 
\left(\frac{r}{\ell_{0}}\right)^{m} 
\exp{\left(\frac{-r^{2}}{4\ell_{0}^{2}}\right)}e^{im\theta} \; ,
\end{equation}
The full one-body wavefunction  
corresponding to $j$th subband may be written 

\begin{equation}
\Phi_{j,m}(r,\theta,z)=\varphi_{m}(r,\theta)\psi_{j}(z) \; ,
\end{equation}
with the corresponding charge distribution function

\begin{equation}
n({\bf x})=\sum_{j}\sum_{m}|\varphi_{m}(r,\theta)|^{2}|\psi_{j}(z)|^{2} \; .
\label{num-1}
\end{equation}

In the strong magnetic field regime, 
where all electrons are accomodated within the LLL 
and execute cyclotron orbits with a common kinetic energy,
the summation over $m$ yields $\frac{1}{2\pi\ell_{0}^2}$ independent of $r$,
the in-plain position of electrons.
In the LLL where the filling factor is a fractional number between zero
and one, only the first subband is filled, and the ground state can be described
as an isospin ferromagnet, i.e., a phase coherent state \cite{Kyang94}.
In zero magnetic field under typical experimental circumstances 
\cite{Suen91,Suen92}, the
second subband is partially filled depending on areal density.
In a sufficiently strong magnetic field, the Landau level degeneracy is
high enough that electrons which were in second subband in zero magnetic
field, will be located in the first subband.
The density distribution function in Eq.($\ref{num-1}$) then reduces to

\begin{equation}
n(z)=\frac{\nu}{2\pi\ell_{0}^2}|\psi_{0}(z)|^2 \; .
\end{equation}

Following a procedure similar to that used for ${\rm GaAs/Al_{x}Ga_{1-x}As}$
heterojunctions at zero magnetic 
field \cite{stern84}, we solve for the quantized energy level of $j$th subband, 
$E_{j}$ and its corresponding
envelope function $\psi_{j}(z)$ satisfing the following 
Schr$\ddot{\rm o}$dinger equation

\begin{equation}
[\frac{-\hbar^2}{2m^*}\frac{d^2}{dz^2}+V_{\rm eff}(z)]
\psi_{j}(z)=E_{j}\psi_{j}(z) \; ,
\end{equation}
Here $m^{*}$ is the electron effective mass of GaAs and $V_{\rm eff}(z)$ 
is the one electron effective 
potential of Kohn-Sham local density functional theory (LDFT)
which splits into three different contributions

\begin{equation}
V_{\rm eff}(z)=V_{b}(z)+V_{h}(z)+V^{\uparrow\uparrow}_{xc}(z) \; .
\end{equation}

The quantity $V_{b}(z)$ is the built in potential due to the infinite barrier 
and $V_{h}(z)$ is the
Hartree term due to all coulomb interactions between electrons in the
presence of a uniform density of background ions

\begin{equation}
V_{h}(z)=-\frac{2\pi e^2}{\epsilon}\int d z^\prime 
| z - z^\prime | n(z^\prime) \; .
\end{equation}
$V^{\uparrow\uparrow}_{xc}(z)$ is the exchange-correlation potential 
in the Kohn-Sham local density functional approximation. 
The LDFA exchange-correlation energy, which has been 
studied by Vignale and Rasolt \cite{vignale87}, 
is a functional of the scalar charge density and 
vector current density. In our approximation, 
we ignore the current density term in the exchange-correlation energy and
we assume that all electrons are spin polarized along the magnetic field.
The best functional for our purposes is the
Vosko-Wilk-Nusair exchange-correlation energy, 
$V^{\uparrow\uparrow}_{xc}(z)$, which is parametrising the Ceperly-Alder 
Monte Carlo calculation \cite{Ceperly}, and which
has been applied to three dimensional 
itinerant ferromagnets \cite{Vosko80}. 

Results of our numerical calculations for $\Delta_{SAS}$ for 
$\nu=1$ filling factor and several $N_s$ are 
listed in Table I. In this Table the parameter d, which we define as the 
distance between the peaks in the charge-density profile as shown in 
Fig.(\ref{fig3}), 
decreases as $N_s$ is lowered. The values of $N_s$ which we used 
have been taken from Suen {\em et al.'s} experimental data \cite{Suen92}.
At this point, we may compare the results of the strong magnetic field 
LDFA and calculation of the renormalized $\Delta^B_{SAS}$ of the many body 
Hamiltonian Eq. (\ref{2.1_1}), in which the tunneling term is enhanced
by electron-electron interactions. 
We will evaluate the integral in
Eq.(\ref{2.1_1}) by knowing the $V_{0z}({\bf q})$ which has been
defined in Eq.(\ref{1.8}). We use the zero magnetic field
orthonormal wave functions of the first and second subband to
find the $V_{0z}({\bf q})$ and hence $\Delta^B_{SAS}$.

We used the self-consistent LDFA symmetric-antisymmetric wave 
functions for zero magnetic field to calculate $\Delta^B_{SAS}$ numerically.
The form of exchange-correlation potential which we chose is 
Hedin and Lundqvist \cite{HL} which has been used in an investigation of
${\rm GaAs/Al_xGa_{1-x}As}$ hetero junctions 
in zero magnetic field \cite{stern84}.

One may define the relative difference between the bare and renormalized  
tunneling term, $\delta=\frac{\Delta_{SAS}-\Delta^B_{SAS}}{\Delta_{SAS}}$ 
which is a quantity to measure the deviation of SWQW's charge distribution 
due to the strong magnetic field. 
$\delta$ may calculated by the integral of $V_{0z}({\bf q}) e^{-q^2/2}$ 
which is shown for different densities in Fig.(\ref{fig1}). 
The result of integration is positive (negative) at high (low) densities
and depends on the long wavelength oscillations of $V_{0z}({\bf q})$. 
Our numerical calculation shows that $\delta$ is
small and even negative at low densities and becomes positive with
increasing the density. 
For example, at $N_{s}=0.5\times 10^{11} {\rm cm}^{-2}$,
we find $\delta= -0.1$ which is relatively small due to the long
wavelength cancellation of $V_{0z}({\bf q})$ upon integration, 
and therefore the electron density $n(z)$ is not affected enormously
by the strong magnetic field.
Note that the sign of $\delta$ is negative,
which means that the tunneling term is increased by the strong magnetic field.
In Fig.(\ref{fig2}), $V_{0z}({\bf q})$ is plotted for 
$N_{s}=0.5\times 10^{11} {\rm cm}^{-2}$ and $W=680 (\AA)$.
In Fig.(\ref{fig2}) we compare the electron density of a SWQW in zero
and strong magnetic field by LDFA calculations.
This comparison for $N_{s}=0.5\times 10^{11} {\rm cm}^{-2}$ and $W=680 (\AA)$
shows that the two densities are very close which is indicates qualitative
agreement between the LDFA and many body calculations.

At very large densities $V_{0z}({\bf q})$ is a smoothly 
negative but close to zero and 
hence $\delta$ is an small negative number. It is important to note that
$\delta$ is very small in the DQW regime which is equivalant to large SWQW 
densities and is strictly zero for the true DQW case. 
In the domain of densities which is of interest for us,
both calculation of LDFA and the many body Hamiltonian show a 
decrease of $\Delta_{SAS}$ due to the
strong magnetic field on the order of 1 degree of Kelvin. 
We conclude that the effect of strong magnetic field on the electron
density $n(z)$ is negligible at very low and very high areal electron density.
At intermediate densities, the effect of the strong magnetic field is 
significant and the electron distribution along the $\hat{z}$-direction 
is changed. One ought to be able to measure this effect experimentally.

In Fig.(\ref{fig3}) we plot the result of a
self-consistent LDFA calculation at strong magnetic field for 
$N_s=3.1\times 10^{11}{\rm cm}^{-2}$ where the difference of charge density
in zero magetic field and strong magnetic field is relatively large.
In strong magnetic field, the charge density is proportional to square of 
the lowest subband wave function. 
Therefore the result of nonzero parameter $d$ is due to tunneling 
effect through $V_{\rm eff}(z)$ barrier where the
eigenenergy is less than the top of effective potential. 
Obviously in the presence of strong perpendicular magnetic field,
$d$ is a monotonicaly decreasing function of $N_{s}$,
and at a critical density $N_{sc}$ and critical width $W_c$,
the parameter $d$ continously goes to zero.
For example our numerical calculation shows this transition between two-layer 
and one-layer quantum wells occurs 
at $N_{s}\equiv N_{sc}=0.5\times10^{11}{\rm cm}^{-2}$ 
for a SWQW's width of $680 (\AA)$.
To the best of our knowledge, there is no exprimental results
to report the value of $\Delta^B_{SAS}$ and $d$ \cite{Mansour}
to check with our numerical results.

\section{collective modes excitations of SWQWs fractional quantum Hall state}
\label{sec5}

Recently, there has been much interest in experimental work in DQWs and SWQWs
\cite{Suen91,Boebinger90}, a remarkable effect is observed, namely the
absence of certain IQHE at odd $\nu$ for sufficiently small $\Delta_{SAS}$
and large interalayer distance ($d$). 
These experimental observations have been explained in
a phase diagram proposed by MacDonald {\em et al.}~\cite{Mac90}
for the presence or absence of the IQHE for different DQW parameters, associated 
with the loss of the isomagnon excitation gap.
We generalize their model to SWQWs with the help of experimental parameters
used by Suen {\em et al.}~\cite{Suen92}.
In the previous sections we found that in the presence of
strong prependicular magnetic field
only the first subband is occupied where all electrons are in the isospin state,
$up$. One may expect that the low lying excitations are isospin waves 
(isomagnons) corresponding to a single flipped isospin which propagates
through the system. Therefore we may describe the isomagnons as a bound state
of an electron of one isospin with a hole of opposite isospin.
The Hamiltonian describing the low-energy excitation of the SWQW
system has been obtained in the preceding sections

\begin{equation}
H=-\frac{\Delta^B_{SAS}}{2} \; \sigma_z({\bf q}=0)
+ \frac{1}{2} \int \frac{d^2 {\bf q}}{(2 \pi)^2}
\; \overline{\sigma}_\mu(-{\bf q}) \; V^{\mu\nu} ({\bf q}) 
\; \overline{\sigma}_\nu({\bf q}) \; ,
\label{Ham}
\end{equation}
where the isopotentials $V^{\mu \nu}({\bf q})$ are defined by Eq.(\ref{1.8}).
The calculation of the isomagnon collective-mode energy associated with
the first to second subband excitation, which is a type of
isospin wave in isospin space, is based on the single mode 
approximation \cite{Mac90} and
the free boson model of Holstein-Primakoff transformation which has been 
exploited in the study of ferromagnetic spin waves problem \cite{Kittel}. 
The normalized isomagnon wave function is given by

\begin{equation}
|\Psi_{-}({\bf k}) \rangle = \frac {e^{\ell_{0}^2 k^2 / 4}}{\sqrt N}
\overline{S}_{-} ({\bf k}) |\Psi_{0} \rangle \, .
\end{equation}

Here $|\Psi_{0} \rangle=\bigotimes_{i} |\uparrow \rangle_{i}$ 
is the fully isospin polarized ground state of the system in which
all electrons are located within the first subband and
$\overline{S}_{\pm} ({\bf k})=\Bigl( \overline{\sigma}_{x}({\bf k}) \pm 
i \overline{\sigma}_{y}({\bf k}) \Bigl) / 2$ are isospin lowering (raising) 
operators.
In the absence of an external perpendicular electeric field where the system is
balanced and the first (second) subbands are symmetric 
(antisymmetric) respectively,
the full Hamiltonian can be separated into terms which 
conserves $S_z^{tot}$ $(H_0)$

\begin{eqnarray}
H_0=T+V_{0}+H^{\prime} \; ,
\label{H1}
\end{eqnarray}
where $T$ is the tunneling term and

\begin{eqnarray}
V_{0} =\frac{1}{2} \int \frac{d^2 {\bf q}}{(2 \pi)^2} \Bigl( \;
V_0 ({\bf q}) \overline{\rho}(-{\bf q}) \overline{\rho}({\bf q}) + 
V_z ({\bf q}) \overline{\sigma}^z(-{\bf q}) \overline{\sigma}^z({\bf q})
+ \nonumber \\
V_{0z} ({\bf q}) [ \; \overline{\sigma}^z(-{\bf q})\overline{\rho}({\bf q})+
\overline{\rho}(-{\bf q})\overline{\sigma}^z({\bf q}) \; ] \; \Bigl ) \; ,
\end{eqnarray}
and a term $H^{\prime}$ which creates (annihilates) 
a pair of isomagnons and hence changes $S_z^{tot}$ by $2$:
\begin{eqnarray}
H^\prime = \frac{1}{2} \int \frac{d^2 {\bf q}}{(2 \pi)^2} 
V_x ({\bf q}) \Bigl( \;
\overline{S}_{+}(-{\bf q})\overline{S}_{+}({\bf q}) + 
\overline{S}_{-}(-{\bf q})\overline{S}_{-}({\bf q}) \; \Bigl) \; .
\label{H2}
\end{eqnarray}

Using the single mode approximation \cite{GMP86} and commutation
relation Eq.(\ref{2.10}) yields

\begin{eqnarray}
\langle \Psi_{-}({\bf k}) | H_0 | \Psi_{-}({\bf k}) 
\rangle = E_0 + \varepsilon({\bf k}) \; .
\end{eqnarray}
Here $E_0=\langle \Psi_0 | H_0 | \Psi_0 \rangle$ and

\begin{eqnarray}
\noindent \varepsilon({\bf k}) = \Delta^B_{SAS}+
\frac{\nu}{2\pi\ell_0^2} V_x({\bf k}) e^{-\ell_0^2 k^2/2} 
-\int \frac{d^2 {\bf q} e^{-\ell_0^2 q^2/2}}{(2 \pi)^2} 
V_{0z}({\bf q}) \Bigl( \; \tilde{h}({\bf q}) + 1 \; \Bigl) 
\nonumber \\
+\int \frac{d^2 {\bf q} e^{-\ell_0^2 q^2/2}}{(2 \pi)^2} \Bigl( \;
V_x({\bf q}) \tilde{h}({\bf q}+{\bf k}) - [\; V_0({\bf q})+
V_z({\bf q})\; ]\tilde{h}({\bf q})\; \Bigl) \nonumber \\
+\int \frac{d^2 {\bf q} e^{-\ell_0^2 q^2/2}}{(2 \pi)^2} \Bigl( \;
[\; V_0({\bf q})+V_z({\bf q})\; ]\tilde{h}({\bf q})
e^{i \ell_0^2 {\bf q}.(\hat{z}\times {\bf k})}\; \Bigl) \; ,
\label{Iso1}
\end{eqnarray}
where $\tilde{h}({\bf q})=h({\bf q}) \exp(\ell_0^2 q^2/2)$ and 
$h({\bf q})$ is the Fourier transform of the pair-correlation 
function. 
For $\nu=1$ this is given by $h({\bf q})= -\exp(-\ell_0^2 q^2/2)$. 
These results generalize the results which were 
obtained previously \cite{Mac90} for the special case
$V_z({\bf q})=V_{0z}({\bf q}) \equiv 0$ for DQWs. 
The physical interpretation of Eq.(\ref{Iso1}) has been explained
in the study of DQW system by MacDonald {\em et al.}~\cite{Mac90}.
The first term of Eq.(\ref{Iso1})
is the dressed excitation energy from the first to the second subband,
the second term is the Hartree correction due to the 
isospin dependent part of the electron-electron
interaction and the third term is the Hartree enhancement of $\Delta^B_{SAS}$ 
associated with isospin coulomb interaction and is zero for $\nu=1$.
The fourth term is the difference between the self-energy of 
electrons in the second and first subbands and the last term is the energy
associated with the electron-hole interaction.

The full Hamiltonian Eq.(\ref{Ham}) may be approximated by a 
soluble quadratic bosonic effective Hamiltonian by 
application of the Holstein-Primakoff transformation

\begin{equation}
H_{b}=E_0 + \sum_{{\bf k}} \Bigl( \;
\varepsilon({\bf k}) \; \hat{b}^{\dagger}_{{\bf k}}\hat{b}_{{\bf k}} +
\frac{1}{2} \lambda({\bf k}) \; (\hat{b}_{{\bf k}}\hat{b}_{-{\bf k}}
+\hat{b}^{\dagger}_{{\bf k}}\hat{b}^{\dagger}_{-{\bf k}}) \; \Bigl) \; ,
\label{Boson}
\end{equation}
where $\hat{b}^{\dagger}_{{\bf k}} (\hat{b}_{{\bf k}})$ 
are bosonic creation (annihilation) operators.
Here $H_0$ and $H^\prime$ are transformed to quadratic boson pieces 
$\sum_{{\bf k}}\varepsilon({\bf k}) \; 
\hat{b}^{\dagger}_{{\bf k}}\hat{b}_{{\bf k}}$ and  
$\frac{1}{2} \sum_{{\bf k}} \lambda({\bf k}) \; 
(\hat{b}_{{\bf k}}\hat{b}_{-{\bf k}}
+\hat{b}^{\dagger}_{{\bf k}}\hat{b}^{\dagger}_{-{\bf k}})$. 
The latter
creates or destroys an isomagnon pair with opposite momentum
and is responsible for broken global gauge symmetry.

One may obtain $\lambda({\bf k})$ by fitting to the exact expression for
the matrix element of $H^\prime$ between the zero and two isomagnon states, 
which leads to 

\begin{eqnarray}
\lambda({\bf k})=\frac{\nu}{2\pi\ell_0^2} V_x({\bf k}) e^{-\ell_0^2 k^2/2}
+\int \frac{d^2 {\bf q} e^{-\ell_0^2 q^2/2}}{(2 \pi)^2}  \;
V_x({\bf q})\tilde{h}({\bf q}+{\bf k}) 
e^{i \ell_0^2 {\bf q}.(\hat{z}\times {\bf k})} \; .
\label{L1} 
\end{eqnarray}

The effective Hamiltonian $H_b$ may be diagonalized by 
Bogoliubov transformation and
the resulting isospin-excitation energies are given by

\begin{equation}
E({\bf k})=\sqrt{\varepsilon^2({\bf k})-\lambda^2({\bf k})} \; .
\label{3.1}
\end{equation}
The energy of the system is reduced by generating virtual isomagnon-pairs and 
therefore, pairing of isomagnons is prefered for the ground state. 
At $\nu=1$ the dispersion relation Eq.(\ref{3.1}) has a gap at ${\bf k}=0$
corresponding to $\Delta^B_{SAS}$ where the quantum Hall gap is specified
by the local minimum of Eq.(\ref{3.1}). For zero tunneling,
Eq.(\ref{3.1}) describes a gapless goldestone mode corresponding to
spontaneous global U(1) symmerty breaking in the ground state
describing a superfluid
associated with fluctuation in $S^{tot}_z$ \cite{Wen92}, in spite of 
violating the U(1) symmetry of the original Hamiltonian Eq.(\ref{Ham})
where the symmetry group of SWQWs Hamiltonian is $Z_2$
corresponding to $\sigma^y \rightarrow -\sigma^y$, while
the DQWs Hamiltonian has U(1) symmetry group when 
$V_{z}({\bf q})=V_{0z}({\bf q})=0$. Note that $Z_2$ symmetry of Hamiltonian
is also broken in the presence of the tilted magnetic field as we mentioned
before.
However at $\nu \neq 1$ where the third term of Eq.(\ref{Iso1}) 
is nonzero but constant (depending on total filling factor), 
the collective modes are gapful even at zero tunneling term and
there is no goldestone mode corresponding to spontaneous symmetry breaking
while in DQW, spontaneous global U(1) symmetry breaking occurs at any filling 
factor $\nu$, generates the goldestone modes for zero tunneling.
The collective modes are understood by
the pole of isospin density-density response function within LLL which
has been called isomagnons. In our approximation it
corresponds to the local minimum of dispersion curve
Fig.(\ref{fig5}), which are gapful or gapless 
in special circumstances as we mentioned above.
Note that this low-energy mode is the analogous to those in the 
Feynman theory of superfluid $^4$He \cite{PG}.
One may associate the odd total filling factor quantum Hall effect
with isomagnons gap where $\Delta^B_{SAS}$,
the energy difference between the subbands is the the 
relevant energy scale to compare with
the in-plane coulomb energy $e^2/(\epsilon_0 \ell_0)$, i.e.
the competition between $\Delta^B_{SAS}$ and the coulomb energy
yields the correspondig QHE. Therefore 
the phase boundary between incompressible and compressible states can be
defined by the collapse of the gap which occurs at certain $\Delta^B_{SAS}$ 
and $d$, while the corresponding gap at even total filling factors has
been identified by the charge excitations (magnetorotons) which is 
infinitely large in our approximation \cite{PG}.
One may generalize the above formalism for FQHE associated with
a many body energy gap due to correlation between electrons in different 
layers which is destroyed if they are uncorrelated by increasing the
layer distance.

At this point we may study the effect of applying a perpendicular 
electric field on the dispersion relation of collective modes
by considering an imbalanced SWQW. As we disscused above,
inversion symmetry is broken by perpendicular electric field and
the first (second) subbands are no longer symmetric (antisymmetric).
There is also an additional term in collective mode 
Hamiltonian, Eq.(\ref{H1}-\ref{H2}),
due to the external electric field Hamiltonian Eq.(\ref{1.12}).
We have to mention that, the corresponding additional term, $V_U$, 
changes $S_z^{tot}$ by 1. In this case the charge excitations couples
with isomagnons and the ground state may be described by the virtual
isomagnon-plasmon pairs. One may divide $V_U$ into two parts, 
$V_U=V_U^0+V_U^z$, where $V_U^0$ ($V_U^z$) 
can also be separated into two parts

\begin{mathletters}
\label{6.1}
\begin{equation}
V_U^0=V_U^{+(0)} \; + \; V_U^{-(0)} \; ,
\label{6.1a}
\end{equation}
\begin{equation}
V_U^z=V_U^{+(z)} \; + \; V_U^{-(z)} \; ,
\label{6.1b}
\end{equation}
\end{mathletters}
where we define $V_{U}^{\pm}$ by

\begin{mathletters}
\label{6.2}
\begin{equation}
V_U^{\pm(0)}=\frac{1}{2} \int \frac{d^2 {\bf q}}{(2\pi)^2} \; 
V_{0x}({\bf q}) \; [ \; 
\overline{\rho}(-{\bf q})\overline{S}_{\pm}({\bf q}) +
\overline{S}_{\pm}(-{\bf q})\overline{\rho}({\bf q}) \; ] \; ,
\end{equation}
\begin{equation}
V_U^{\pm(z)}=\frac{1}{2} \int \frac{d^2 {\bf q}}{(2\pi)^2} \; 
V_{xz}({\bf q}) \; [ \; 
\overline{S^z}(-{\bf q})\overline{S}_{\pm}({\bf q}) +
\overline{S}_{\pm}(-{\bf q})\overline{S^z}({\bf q}) \; ] \; .
\end{equation}
\end{mathletters}

Again $V_U^0$ ($V_U^z$) will be bosonized by the
Holestein-Primakoff transformation

\begin{mathletters}
\label{6.3}
\begin{equation}
H_b^0=\frac{1}{2} \sum_{{\bf k}} \; \eta({\bf k}) \; \Bigl(
\hat{a}_{{\bf k}} \hat{b}_{-{\bf k}} + 
\hat{b}^\dagger_{-{\bf k}}\hat{a}^\dagger_{{\bf k}}  \; \Bigl) \; ,
\label{6.3a}
\end{equation}
\begin{equation}
H_b^z=\frac{1}{2} \sum_{{\bf k}} \; \zeta({\bf k}) \; \Bigl(
(\hat{n}_{{\bf k}}-\frac{1}{2})\hat{b}_{-{\bf k}} + 
\hat{b}^\dagger_{-{\bf k}}(\hat{n}_{{\bf k}}-\frac{1}{2})  \; \Bigl) \; ,
\label{6.3b}
\end{equation}
\end{mathletters}
where we have used the bosonic hard sphere condition \cite{Fradkin} for 
Eq.(\ref{6.3b}). Here $\hat{a}^\dagger_{{\bf k}} (\hat{a}_{{\bf k}})$ 
is the charge density excitation creation (annihilation) operator,
which yield no pole in the density-density response function within
LLL \cite{Fertig89}, e.g. magnetorotons
with a huge gap equivalent to $\hbar \omega_c$ and
$\hat{n}_{{\bf k}}$ which is bosonic number operator.
Therefore, Eq.($\ref{6.3}$) may describe coupling between isomagnons 
and magnetorotons due to external electric field bias.

Following the same calculation we have done for the ballanced SWQW 
Eq.(\ref{L1}), we may determine $\eta({\bf k})$ $\Bigl(\zeta({\bf k})\Bigl)$
from the exact result for matrix element of $V_U^0$ ($V_U^z$) between
the zero and two isomagnon-magnetoroton states 

\begin{eqnarray}
\eta({\bf k}) = \int \frac{d^2 {\bf q}e^{-\ell_0^2 q^2/2}}{(2\pi)^2} 
V_{0x}({\bf q}) \Big(
1+\tilde{h}({\bf q})+\tilde{h}({\bf k})+
e^{i \ell_0^2 {\bf q}.(\hat{z}\times{\bf k})}
\tilde{h}({\bf k}+{\bf q}) \; \nonumber \\ 
+ \frac{e^{ (q\overline{k}-k^2)/2}}{2} 
[ \; s(-{\bf k},-{\bf q})+s({\bf k},{\bf q})-2 \; ] \Big) \; ,
\end{eqnarray}

\begin{eqnarray}
\zeta({\bf k}) = \int \frac{d^2 {\bf q}e^{-\ell_0^2 q^2/2}}{(2\pi)^2} 
V_{xz}({\bf q}) \Big(
\frac{1}{2}+\tilde{h}({\bf q})+\tilde{h}({\bf k})+
e^{i \ell_0^2 {\bf q}.(\hat{z}\times{\bf k})}
\tilde{h}({\bf k}+{\bf q}) \; \nonumber \\ 
+ \frac{e^{ (q\overline{k}-k^2)/2}}{2} 
[ \; s(-{\bf k},-{\bf q})+s({\bf k},{\bf q})-2 \; ] \Big) \; ,
\end{eqnarray}

\noindent where $s({\bf k}_1,{\bf k}_2)$ is the 
Fourier transform of three-point correltion function

\begin{equation}
s({\bf k}_1,{\bf k}_2)=1+\frac{1}{N} \sum_{i} \sum_{j\neq i} 
\sum_{n\neq j\neq i} 
\langle B_i({\bf k}_1) B_j({\bf k}_2) B_n(-{\bf k}_1-{\bf k}_2) \rangle_0\; ,
\end{equation}
and $B_j({\bf k}) \equiv \overline{e^{i{\bf k}.{\bf r}_j}} =
e^{-\frac{\vert k\vert^2}{4}}\> \tau_k(j) $, which is projected 
free electron wave function onto LLL has been defined by 
Eq.(\ref{eq2.4}-\ref{eq3.90}) and $\langle ... \rangle_0$ is the
expectation value with respect to ground state many body wave function,
$| \Psi_0 \rangle$. 
It is possible to find analytical 
expression of three-point correltion function for $\nu=1$ quantum Hall
state where the many body ground state wave function
may be written as a single Slater determinant \cite{JMG}. The 
three-point correlation function in real space is 
\begin{equation}
g({\bf r}_1, {\bf r}_2, {\bf r}_3) = 
1-e^{-\frac{|z_2-z_1|^2}{2}}-e^{-\frac{|z_3-z_2|^2}{2}}
-e^{-\frac{|z_1-z_3|^2}{2}} + 2 e^{\frac{U}{2}} \cos \frac{V}{2} \; ,
\end{equation}
where $z_j = x_j + i y_j$ is in-plane complex coordinate of particles and
\begin{eqnarray}
&U=-(r^2_1+r^2_2+r^2_3) + {\bf r}_1.{\bf r}_2 + {\bf r}_2.{\bf r}_3
+ {\bf r}_1.{\bf r}_3 \; , \nonumber \\ &
V={\bf r}_3 \wedge {\bf r}_2 + {\bf r}_2 \wedge {\bf r}_1
+ {\bf r}_1 \wedge {\bf r}_3 \; ,
\end{eqnarray}
with ${\bf r}_j \wedge {\bf r}_k = x_j y_k - y_j x_k$.

We can calculate the isomagnon dispersion relation Eq.(\ref{3.1}) 
numerically after calculating realistic form-factors for SWQWs by 
using realistic first and second subband wave functions.
We used LDFA techniques with Hedin and Lundquvist exchange-correlation
energy functional in zero-magnetic field to obtain the subband
wave functions. 
Therefore we may obtain the phase diagram for
the $\nu=1$ filling factor in a SWQW structure in 
the relevant $d-\Delta_{SAS}$ parameter space. 

In order to make a comparison with the results for DQWs \cite{Mac90},
we associate the vanishing of the isomagnon excitation gap with 
the loss of incompressiblity of quantum Hall states
to obtain a phase boundary for missing the odd integral QHE,
as it has been defined for DQWs by MacDonald 
{\em et al.}~\cite{Mac90}. In Fig.(\ref{fig5}) and Fig.(\ref{fig6}) we
show the results of our calculation at $\nu=1$ for the 
isomagnon dispersion relation and the
phase boundary for SWQW with different densities. 
At large $d$ a SWQWs may be described effectively by a DQW.
In this limit the phase boundary of the SWQW 
is the same as a DQW with the appropriate parameters.
The main difference between the DQW and SWQW phase boundaries appears
when $d$ is small, which correspond to small width or density.
In contrast to DQW,
in SWQWs phase boundary, $d$ goes to zero where $\Delta_{SAS}$ is finite,
$\Delta^c_{SAS}=\Delta_{SAS}(d=0)$, due to critical charge density effect
where the short range component of the $isopin$ coulomb interaction
is soften and the system evolve to one-component normal state.
In Fig.(\ref{fig6}), our phase boundary is compared with DQWs and
the result of LDFA in zero magnetic field for a given well width.
The latter is consistent with the reported phase transition \cite{Suen91}.
As Fig.(\ref{fig6}) shows, the phase transition occurs at smaller
$\Delta_{SAS}$ in comparison with the result of
MacDonald {\em et al.} \cite{Mac90} for a given $d$. 
These results illustrate that our model leads to a 
more realistic prediction
for phase trnsition in comparison with the previous models, however, 
our results can not capture the whole experimental datas which has been
reported for phase boundary \cite{Suen91,Suen92}.
Note that the experimental results of Princeton group 
\cite{Suen91,Suen92} are based on the zero magnetic field 
measurements of $\Delta_{SAS}$ and calculated $d$ by LDFA.

\section{conclusion}
We have presented the study of the quantum Hall effect in
the single wide quantum wells. We showed that the effect of
integral overlap in the middle of the SWQW has a significant effect
to the quantum Hall states in these systems.

We showed that the tunneling term ($\Delta_{SAS}$) and the
interlayer separation ($d$) are both renormalised due to the 
strong magnetic field. One may observe these effects experimentally.
We also used and comparing different techniques, 
projected many body Hamiltonian onto lowest Landau levels,
Hartree-Fock approximation and local density functional approximation
to evaluate the tunneling term and interlayer separation.
Our results confirm that the tunneling term decreases by the strong 
magnetic field and hence increasing the interlayer separation.

We found a phase boundary for integral quantum Hall phase transition 
at $\nu=1$. We showed that our model cover the most part of
phase boundary which has been reportaed experimentally \cite{Suen92}
in comparison with pervious models \cite{Mac90,He-93}.

\acknowledgements
It is pleasure to acknowledge useful conversation with S. M. Girvin,
Song He, A. H. MacDonald, S. Rouhani and M. Shayegan.
The work at Indiana University is supported by grant 
DMR-9416906. One of us (MA) acknowledges The Center for Thoretical
Physics and Mathematics, AEOI, Tehran, Iran for the
financial support.



\begin{figure}
\caption{Numerical calculated electron wavefunction of zero magnetic field
symmetric and antisymmetric subbands of a SWQW with $W=1000 (\AA)$
and $N_s=3.1\times 10^{11} {\rm cm}^{-2}$. The envelop wavefunction 
in the middle
of the sample is small due to tunneling through the self consistent potential 
barrier of the SWQW. The symmetric and antisymmetric wave functions are
close to each other and DQW is a good approximation for this case.
}
\label{fig0}
\end{figure}

\begin{figure}
\caption{$V_{xx}({\bf q})$ (solid line) and $V_{zz}({\bf q})$ (dash line)
are shown for two different SWQW with $W=680 (\AA)$, 
$N_s=1.8\times 10^{11} {\rm cm}^{-2}$ Fig.(a)
and $N_s=3.1\times 10^{11} {\rm cm}^{-2}$ Fig.(b).
The $V_{0z}({\bf q})$ is supressed by increasing the areal electron density.
$V_{zz}$ in (a) is scaled by 10 and in (b) by 100 to show in the above
figures.
}
\label{fig01}
\end{figure}

\begin{figure}
\caption{Numerical results of $V_{0z}({\bf Q}\ell_0)$
vs ${\bf Q}\ell_0$ is shown for three different SWQW's densities with 
$W=680 (\AA)$. 
Varing areal densities $N_s=1.8, 2.3$ and $3.1\times 10^{11} {\rm cm}^2$ 
are shown by long dashed, dashed and solid curves.
The inset exhibits the calculated electron wave function of 
zero magnetic field
symmetric and antisymetric subbands of a SWQW for
$N_s=3.1\times 10^{11} {\rm cm}^{-2}$.
}
\label{fig1}
\end{figure}

\begin{figure}
\caption{Numerical results of $V_{0z}({\bf Q}\ell_0)$
vs ${\bf Q}\ell_0$ is shown for SWQW's electron areal density
$0.5\times 10^{11} {\rm cm}^{-2}$ with $W=680 (\AA)$.
The integral of $V_{0z}({\bf Q}\ell_0)e^{-Q^2\ell_0^2/2}$ is small
and therefore the $\Delta_{SAS}$ enhancement due to strong $B$ is small.
The inset exhibits the calculated electron density along
$z-$direction for two different regimes, zero magnetic field
(dotted points) and strong magnetic field (dashed line)
which demonstrates a quasi 3D electron system.
These two densities profile are nearly the same and the effect of
strong magnetic field on the charge density along the $z-$axis
is negligible which is true for 3D electron system.
}
\label{fig2}
\end{figure}

\begin{figure}
\caption{Electron density for self-consistent calculations in zero 
(solid curve) and strong magnetic field (dashed curve)
for $N_s=3.1\times 10^{11} {\rm cm}^{-2}$ and 
$W=680 (\AA)$. The effect of strong $B$ is decreasing (increasing) the
probability of finding an electron in the edge (middle) of the SWQW
which means that electrons are pushed to go in the middle of SWQW
by the magnetic field.
}
\label{fig3}
\end{figure}

\begin{figure}
\caption{Calculated phase diagram for the $\nu=1$ QHE in SWQW system.
The lines denote the boundaries of phase corresponding to 
compressible-incompressible transition. These calculation has
done for three differnt SWQW's width 400, 680 and 800 $\AA$.
The phase boundary touch horizontal axis at $\Delta^c_{SAS}$ which
corresponds to critical density $N_{sc}$ and critical width $W_c$.
$\Delta^c_{SAS}$ decreased with increasing of $W$.
This is the crossover between one layer compressible and incompressible
quantum Hall systems.
The solid line is the calculated phase boundary of DQW for
$\delta-$layers separated by a distance $d$ which is the asymptotic
limit of large SWQW's width, i.e. the behavior of phase boundary for 
the sample $W=800$ is closer to the DQW.
The inset exhibits the isomagnon dispersion relation for a SWQWs
width $W=680$ and the areal density $N_{s}=1.8 \times 10^{11}{\rm cm}^{-2}$
at the phase boundary where the isomagnon gap is zero. The energy of
isomagnons at zero wave vector is specified by the $\Delta^B_{SAS}$
where at $\nu=1$, the zero tunneling yields gapless dispersion relation.
}
\label{fig5}
\end{figure}

\begin{figure}
\caption{The comparison between the realistic (triangles) for a SWQW with
$W=680 \AA$ width and DQW (solid line) phase diagrams are shown in this figure.
The results of our zero magnetic local density functional results 
(circles) is shown. The integral quantum Hall phase transition 
has been reported in $\Delta_{SAS}/(e^2/\epsilon_0 \ell_0) \sim 0.05$
at zero magnetic field. The results of our calcuation
is closer to the experimental phase boundary than the results 
of DQW approximation.
}
\label{fig6}
\end{figure}
\begin{table}
\caption{Comparison of zero magnetic field $\Delta_{SAS}$ which is 
obtained by Hedin and Lundqvist LDFA in zero magnetic field 
and renormalized $\Delta^B_{SAS}(\nu=1)$, $\Delta^{\prime B}_{SAS}(\nu=1)$
which is obtained by many body Hamiltonian and LDFA in the presence of
strong perpendicular magnetic field 
and also renormalized $\Delta^B_{SAS}$ with filling factor $\nu=1/2$,
for a 680($\AA$) wide well vs areal density $N_s$.}
\begin{tabular}{lcccc}
$N_{s}(10^{11}{\rm cm}^{-2})$ & $\Delta_{sas} \; (K)$ & 
$\Delta^B_{sas}(\nu=1) \; (K)$ &
$\Delta^{\prime B}_{SAS}(\nu=1) \; (K)$ &
$\Delta^{B}_{SAS}(\nu=1/2) \; (K)$ 
\\ \tableline
1.8&12.5&12.2&11.4&12.1 \\
2.3& 9.5& 8.2& 8.5& 7.7\\
2.8& 7.4& 5.8& 6.6& 5.0\\
3.1& 6.4& 4.7& 5.7& 4.0\\
\end{tabular}
\end{table}

\end{document}